\documentclass[useAMS]{mn2e}
\usepackage{array}
\usepackage{epsfig}
\usepackage{lscape}
\usepackage{footnote}
\usepackage{graphics}
\usepackage{etoolbox}
\usepackage{leftidx}
\usepackage{footnote}
\usepackage{color}
\usepackage{cleveref}
\usepackage{caption}
\usepackage{subfig}
\usepackage{url}

\title[Extended Lyman-$\alpha$ emission around quasar J0823$+$0529]{
A coronagraphic  absorbing cloud reveals the narrow-line region and
extended Lyman-$\alpha$ emission of QSO J0823$+$0529
\thanks{Based  on  data  obtained  with  MagE  at  the
Clay  telescope  of the  Las  Campanas  Observatory
(CNTAC  Prgm. ID CN2012B-51 and CN2013A-121)}}
\author[H. Fathivavsari et al.]
  {H.~Fathivavsari$^{1}$,
  P.~Petitjean$^{1}$, P.~Noterdaeme$^{1}$,
  I. P\^aris$^{2}$, H. Finley$^{3,4}$,
  \newauthor 
   S. L{\'o}pez$^{5}$,
   R. Srianand$^{6}$, P. S{\'a}nchez$^{5}$\\
  $^{1}$Institut d'Astrophysique de Paris, Universit\'e Paris 6-CNRS, UMR7095, 98bis Boulevard Arago, 75014 Paris, France\\
  $^{2}$Osservatorio Astronomico di Trieste, via G. B. Tiepolo 11, 34131 Trieste, Italy\\
  $^{3}$CNRS/IRAP, 14 Avenue E. Belin, F-31400 Toulouse, France\\
  $^{4}$University Paul Sabatier of Toulouse/ UPS-OMP/ IRAP, F-31400 Toulouse, France\\
  $^{5}$Departamento de Astronom\' ia, Universidad de Chile, Casilla 36-D, Santiago, Chile\\
  $^{6}$Inter-University Centre for Astronomy and Astrophysics, Post Bag 4, Ganeshkhind, 411 007, Pune, India\\
}

\begin{document}

\date{Accepted .......   Received ....... }

\pagerange{\pageref{firstpage}--\pageref{lastpage}} \pubyear{2002}

\maketitle

\label{firstpage}

\begin{abstract}
We report long-slit spectroscopic observations of the quasar SDSS
J082303.22+052907.6 ($z_{\rm CIV}$$\sim$3.1875), whose Broad Line
Region (BLR) is partly eclipsed by a strong damped Lyman-$\alpha$
(DLA; log$N$(HI)=21.7) cloud. This allows us to study the Narrow
Line Region (NLR) of the quasar and the Lyman-$\alpha$ emission from
the host galaxy. Using {\sc cloudy} models that explain the presence
of strong NV and PV absorption together with the detection of
SiII$^*$ and OI$^{**}$ absorption in the DLA, we show that the
density and the distance of the cloud to the quasar are in the
ranges 180~$<$~$n_{\rm H}$~$<$~710~cm$^{-3}$ and
580~$>$~$r_0$~$>$230~pc, respectively. Sizes of the
neutral($\sim$2-9pc) and highly ionized phases ($\sim$3-80pc) are
consistent with the partial coverage of the CIV broad line region by
the CIV absorption from the DLA (covering factor of $\sim$0.85). We
show that the residuals are consistent with emission from the NLR
with CIV/Lyman-$\alpha$ ratios varying from 0 to 0.29 through the
profile. Remarkably, we detect extended Lyman-$\alpha$ emission up
to 25kpc to the North and West directions and 15kpc to the South and
East. We interpret the emission as the superposition of strong
emission in the plane of the galaxy up to 10kpc with emission in a
wind of projected velocity $\sim$500km~s$^{-1}$ which is seen up to
25kpc. The low metallicity of the DLA (0.27 solar) argues for at
least part of this gas being in-falling towards the AGN and possibly
being located where accretion from cold streams ends up.
\end{abstract}

\begin{keywords}
galaxies: evolution -- intergalactic medium -- quasars: absorption
lines -- quasars: individual: SDSS J082303.22$+$052907.6
\end{keywords}

\section{Introduction}

Luminous high-redshift quasars consist of supermassive black holes
residing at the center of massive galaxies and growing through mass
accretion of gas in an accretion disk. Bright quasars play an
important role in shaping their host galaxies through the emission
of ionizing flux but also through launching powerful and
high-velocity outflows of gas. These outflows inject energy and
material to the disk of the galaxy and may be influencing the
physical state up to larger distances. It has remained unclear
however what are the mechanisms that drive energy from the very
center of the active galactic nuclei (AGN) to the outskirts of the
galaxy.

Observational evidence for outflows and winds in AGNs is  seen as
prominent radio-jets in radio-loud sources, broad absorption lines
observed in broad absorption line (BAL) quasars, or through the
photoionized warm absorber  which is frequently observed  in the
soft  X-rays  (e.g.  Crenshaw et al. 2003). Gravitational
micro-lensing  studies  have  shown  that  the  primary X-ray
emission  region  in  AGN  is  of  the  order  of  a  few  tens  of
gravitational  radii in  size  (e.g. Dai et al. 2010) and X-ray
spectroscopy shows that highly ionized outflows launched from this
region are seen in high-$z$ quasars (Chartas et al. 2009) and in at
least 40\% of them with velocities up to 0.1~c (Gofford et al.
2013).

Outflows are observed also on large scales. Mullaney et al. (2013)
used the SDSS spectroscopic data base to study the one-dimensional
kinematic properties of [OIII]$\lambda$5007 by performing
multicomponent fitting to the optical emission-line profiles of
about 24000, $z<0.4$ optically selected AGNs. They showed that
approximately 17 percent of the AGNs have emission-line profiles
that indicate their ionized gas kinematics are dominated by outflows
and a considerably larger fraction are likely to host ionized
outflows at lower levels. Harrison et al. (2014) find high-velocity
ionized gas (velocity widths of about 600-1500~km~s$^{-1}$) with
observed spatial extents of (6-16)~kpc in a representative sample of
$z<0.2$, luminous (i.e. $L$[O~{\sc iii}]$>$10$^{41.7}$~erg~s$^{-1}$)
type 2 AGNs. Therefore galaxy-wide energetic outflows are not
confined to the most extreme star-forming galaxies or radio-luminous
AGNs.

If outflows are observed both on small and large scales, how the
small scale outflows are transported at larger distances remains
unclear (Faucher-Gigu\`ere et al. 2012, Ishibashi \& Fabian 2015,
King \& Pounds 2015). This is however a crucial question as these
outflows are massive and energetic enough to significantly influence
star formation in the host galaxy and provide significant metal
enrichment to the interstellar and intergalactic media (e.g. Dubois
et al. 2013). At high redshift where quasars are more luminous, the
consequences of such outflows are of first importance for galaxy
formation. One way to study the interplay between the quasar and its
surrounding is to search for Lyman-$\alpha$ emission around quasars
(e.g. Hu \& Cowie 1987, Hu et al. 1996, Petitjean et al. 1996,
Bunker et al. 2003, Christensen et al. 2006). These observations
reveal gas infalling onto the galaxy (Weidinger et al. 2005),
positive feed-back from the AGN (Rauch et al. 2013) or a correlation
between the luminosity of the extended emission and the ionizing
flux from the quasar (North et al. 2012).

Very recently, we searched quasar spectra from the SDSS-III Baryon
Oscillation Spectroscopic Survey (BOSS; Dawson et al. 2014) for the
rare occurrences where a strong damped Lyman-$\alpha$ absorber (DLA)
blocks the Broad Line Region emission (BLR) from the quasar and acts
as a natural coronagraph to reveal narrow Lyman-$\alpha$ emission
from the host galaxy (Finley et al. 2013; see also Hennawi et al.
2009). This constitutes a new way to have direct access to the
quasar host galaxy and possibly, when the size of the DLA is small
enough, to the very center of the AGN. Out of a total of more than
70,000 spectra of $z>2$ quasars (P\^aris et al. 2012), we gathered a
sample of 57 such quasars and followed-up six of them with the slit
spectrograph Magellan-MagE to search for the very special cases
where the DLA coronagraph reveals the very center of the host galaxy
and extended Lyman-$\alpha$ emission. In the course of this
follow-up program, we found one object SDSS~J0823+0529  where the
DLA does not cover the Lyman-$\alpha$ broad line region entirely and
reveals the emission of the Lyman-$\alpha$ and C~{\sc iv} narrow
line regions. We show here that this is a unique opportunity to
study the link between the properties of the central regions of the
AGN to that of the gas in the halo of the quasar.

The paper is organized as follows. In Section 2 we describe the
observations and data reduction. We derive properties of the gas
associated with the DLA (metallicity, ionization state, density,
distance to the quasar, typical size) in Section~3. We discuss the
properties of the quasar narrow line region and of the extended
Lyman-$\alpha$ emission in Sections 4 and 5, respectively, We then
finally, present our conclusions in Section~6. In this work, we use
a standard CDM cosmology with $\Omega_{\Lambda}$~=~0.73,
$\Omega_{m}$~=~0.27, and H$_0$~=70~km~s$^{-1}$~Mpc$^{-1}$ (Komatsu
et al. 2011). Therefore 1~arcsec corresponds to about 7.1~kpc at the
redshift of the quasar ($z$~=~3.1875 see below). In the following we
will use solar metallicities from Asplund et al. (2009).

\begin{figure*}
\centering
\begin{tabular}{c}
\includegraphics[bb=43 435 554 613,clip=,width=0.90\hsize]{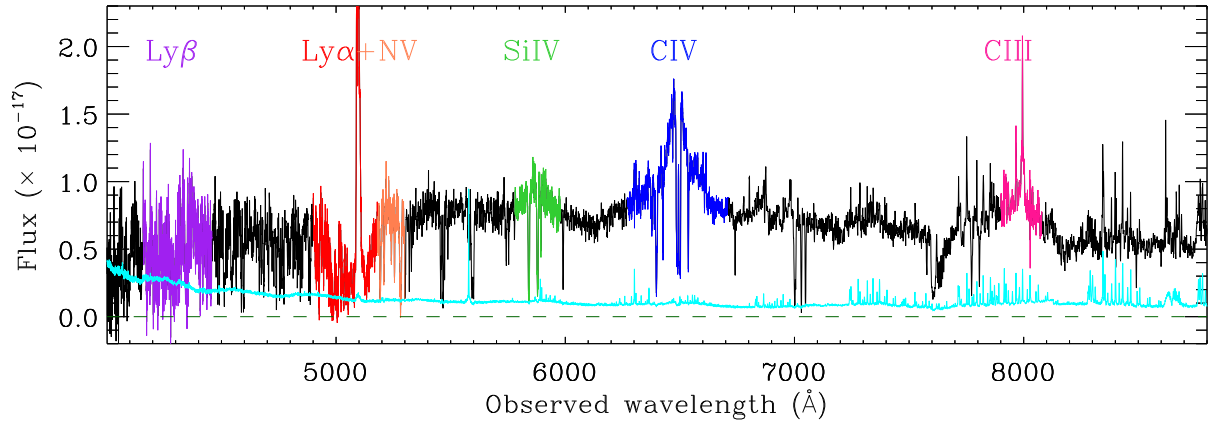}
\end{tabular}
\caption{Flux-calibrated spectrum of the quasar SDSS~J0823+0529 over
the full observed wavelength range. The positions of broad emission
lines are shown with different colors. The Lyman-$\alpha$-N~{\sc v}
broad emission is absorbed by a strong DLA and a strong narrow
Lyman-$\alpha$ emission is revealed. The cyan curve shows the
corresponding error array. The y-axis is in
erg~s$^{-1}$~cm$^{-2}~$\textup{\AA}$^{-1}$.}
 \label{QSO_whole_spect}
\end{figure*}

\begin{figure*}
\centering
\begin{tabular}{c}
\includegraphics[bb=59 353 549 713,clip=,width=0.95\hsize]{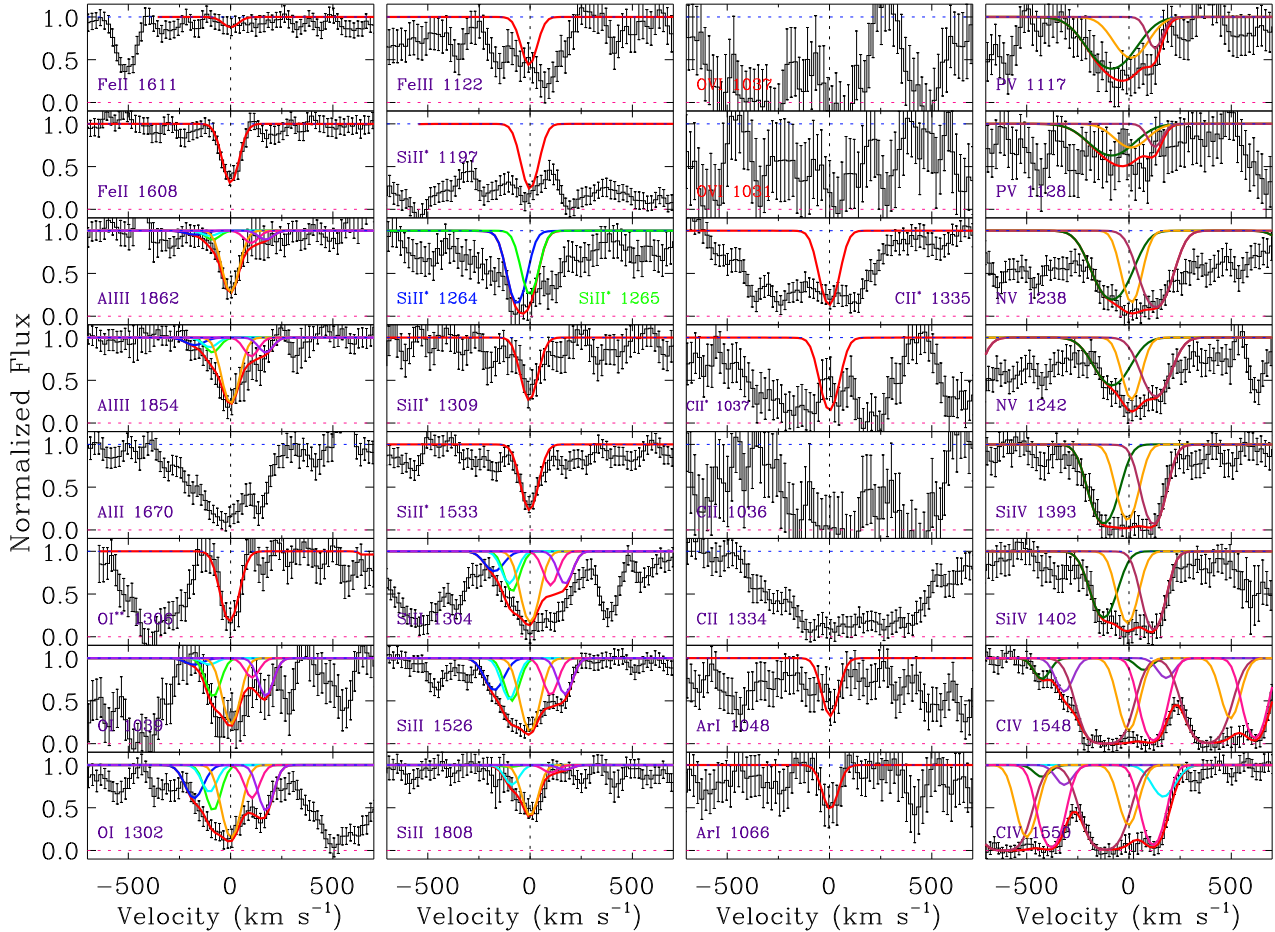}
\end{tabular}
\caption{Absorption profiles detected in the DLA at $z_{\rm
abs}$~=~3.1910. This redshift is used as the origin of the velocity
scale. Multi component Voigt profile fits were performed on the
absorption profiles and the results are shown by red lines together
with individual components plotted in different colors. Single
component fits were performed on the O~{\sc i}$^{**}$, Si~{\sc
ii}$^{*}$, and Ar~{\sc i} absorption lines. Regions corresponding to
the Al~{\sc ii}, C~{\sc ii}, and O~{\sc vi} are not fitted due to
the contaminations by some unidentified absorption lines. }
\label{j08231D}
\end{figure*}

\section{Observations and data reduction}

The spectrum of the quasar SDSS~J0823$+$0529 was observed with the
Magellan Echellete spectrograph (MagE; Marshall et al. 2008) mounted
on the 6.5~m Clay telescope located at Las Campanas Observatory.
MagE is a medium-resolution long-slit spectrograph that covers the
full wavelength range of the visible spectrum
(3200~$\textup{\AA}$~$-$~1~$\mu$~m). Its 10~arcsec long slit and
0.30$^{"}$ per pixel sampling in the spatial direction, are ideal
for observing high-redshift extended astrophysical objects. The
spectrograph was designed to have high throughput in the blue, with
the peak throughput reaching 22~$\%$ at 5600~$\textup{\AA}$
including the telescope.

The quasar was observed in December 2012 with an one arcsec width
slit aligned along two different position angles (PAs) for 1~hour
each. One position was South-North (PA~=~0) and the other position
was East-West (PA~=~90). Another 1-hour exposure with PA~=~90~ was
taken in February 2013 but the resultant spectrum has a lower SNR.
Following each exposure, the spectrum of a standard star was also
recorded allowing us to precisely flux-calibrate the quasar spectra.
The seeing, measured on the extracted trace of the quasar is
1.06~arcsec for PA~=~0 and 0.93~arcsec for PA~=~90.

We reduced the spectrum using the Mage\_Reduce pipeline written in
the Interactive Data Language (IDL) by George Becker\footnote{
\url{ftp://ftp.ast.cam.ac.uk/pub/gdb/mage_reduce/}}. In addition to
the 1-dimensional (1D) spectrum, the pipeline provides 2-dimensional
(2D) sky-subtracted science frames as well as the corresponding 2D
wavelength solution corrected for the vacuum heliocentric velocity
shifts. These 2D images will later be used to rectify the curved
spectral orders (see below). Here, we note that since there is an
extended Lyman-$\alpha$ emission in order 9 of our 2D spectra, we
preferentially used a wider extraction window to extract this order.

Two-dimensional spectra when imaged onto a detector are often curved
with respect to the natural (x,y)-coordinate system of the detector
defined by the CCD columns (Kelson 2003). To rectify the curvature
of the orders we first rebin the wavelength area (given by the 2D
wavelength solution) and position area (given by the 2D slitgrid
array provided by the pipeline) increasing the number of pixel by a
factor of hundred. We define a new 2D array, with one dimension
representing the wavelength ($\lambda$) and the other the position
on the slit ($s$). For each pixel on the 2D spectrum of the quasar,
we take its corresponding wavelength and its position on the slit
directly in the rebinned 2D areas. Therefore, the (x,y) coordinates
of each pixel (defined by the CCD columns) can now be transformed to
an ($s$, $\lambda$) coordinate defined by the new 2D array
introduced above. In this new ($s$, $\lambda$)-coordinate system,
the curvature of the orders and the tilting of the spectral lines
are all rectified. In this study, we use these rectified 2D images
when we discuss the spatial extension of the Lyman-$\alpha$ emission
line in the quasar spectrum. Note that the rebinning process has
very little effect on the Lyman-$\alpha$ emission because it is
conveniently placed in the middle of the corresponding order where
curvature is at minimum.

Finally, the extracted 1D spectrum of each individual order is
corrected for the relative spectral response of the instrument and
flux-calibrated using the spectrum of a standard star (HR~1544)
observed during the same night. The spectrum of the standard star
can be found in the ESO standard star catalogue webpage\footnote{
\url{https://www.eso.org/sci/observing/tools/standards.html}}.  We
emphasize that our standard star spectrum was obtained immediately
after observing the quasar spectrum. We also flux calibrated in the
same way the 2D spectra in the Lyman-$\alpha$ range. These
flux-calibrated spectra are then combined weighting each pixel by
the inverse of its variance. The resulting spectrum (after binning
with a 3~$\times$~3 box) has $\sim$~27~km~s$^{-1}$ per pixel and 3
pixels per resolution element, and therefore
FWHM~$\sim$~80~km~s$^{-1}$.

We have fitted the quasar C~{\sc iv} emission line with two Gaussian
functions (to mimic the two lines of the doublet) to estimate the
redshift of the quasar. We derive $z_{\rm em}$~=~3.1875. This is
$\sim$330~km~s$^{-1}$ smaller than the redshift of the DLA ($z_{\rm
DLA}$~=~3.1910 derived from the fit of Si~{\sc ii} and Fe~{\sc ii}
absorptions). However, it is well known that redshifts from C~{\sc
iv} are smaller by up to 600~km~s$^{-1}$ compared to redshifts from
narrow forbidden lines (e.g. Hewett \& Wild 2010). Therefore we
cannot be certain that the DLA is redshifted compared to the quasar.
We would need to detect [O\,{\sc iii}] lines redshifted to the
infra-red to have a better estimate of $z_{\rm em}$.

\begin{figure}
\centering
\begin{tabular}{c}
\includegraphics[bb=56 352 317 565,clip=,width=0.95\hsize]{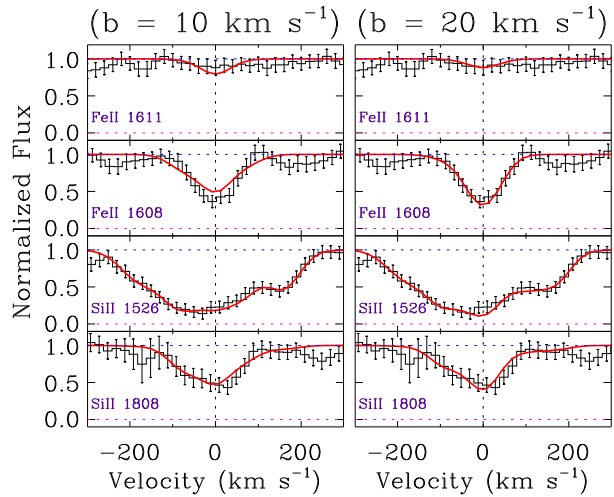}
\end{tabular}
\caption{The relative optical depths of Fe~{\sc ii}$\lambda$1608 and
1611 are used to constrain the Doppler parameter of the main DLA
component. The fit is constrained by Si~{\sc ii} profiles. In each
column, the Doppler parameter of the strong central component is
fixed to the value indicated at the top. }
 \label{dopplerb}
\end{figure}

\section{A DLA acting as a coronagraph}

Figure~\ref{QSO_whole_spect} shows the co-added spectrum of
SDSS~J0823+0529 using all the exposures taken at different PAs. It
is apparent that a strong DLA is located close to the redshift of
the quasar. However, strong Lyman-$\alpha$ emission is present in
the center of the DLA trough. This is possible if the absorbing
cloud has a transverse extension which is smaller than the central
narrow line region (NLR) of the quasar. Note that the strength of
the Lyman-$\alpha$ emission together with the size of the cloud
derived below makes it very improbable that the Lyman-$\alpha$
emission could be a consequence of star formation in the DLA itself.

The redshift of the DLA ($z_{\rm DLA}$~=~3.1910) was determined by
conducting a simultaneous Voigt profile fit of the Fe~{\sc ii},
Si~{\sc ii}$^{*}$, and O~{\sc i}$^{**}$ absorption profiles.

\subsection{Elemental abundances}

A Voigt profile fit was conducted on the damped Lyman-$\alpha$
absorption line of this system, resulting in a neutral hydrogen
column density of log~$N_{\rm HI}$~=~21.70~$\pm$~0.10. Note that the
placement of the quasar continuum is very uncertain as the DLA
covers the Lyman-$\alpha$ and the N~{\sc v}  broad emission lines.
However, the core and especially the red wing of the DLA profile
allowed us to satisfactorily constrain the H~{\sc i} column density.
We will come back to this in Section~3.4.

We detect absorption lines from O~{\sc i}, O~{\sc i}$^{**}$, Si~{\sc
ii}, Si~{\sc ii}$^*$, Fe~{\sc ii}, Al~{\sc ii}, Al~{\sc iii},
Ar~{\sc i}, C~{\sc ii}, C~{\sc ii}$^{*}$, C~{\sc iv}, Si~{\sc iv},
P~{\sc v} and N~{\sc v}. Absorption profiles are shown in
Fig.~\ref{j08231D} and results from fitting these lines are given in
Tables~\ref{lowion}~$\&$~\ref{hiion}. Techniques used here are
similar to those in Fathivavsari et al (2013). It must be noted that
most of the lines are saturated, preventing us from deriving
accurate column densities especially at the resolution of our data
(R$\sim$4000).

The profiles are dominated by a main strong component clearly seen
in particular in Fe~{\sc ii}$\lambda$1608 and Si~{\sc
ii}$^*$$\lambda$1533. To constrain the Doppler parameter of this
component, we take advantage of the fact that Fe~{\sc
ii}$\lambda$1608 is not saturated and well defined while Fe~{\sc
ii}$\lambda$1611 is not detected (see Fig~\ref{dopplerb}). We start
by fitting together Si~{\sc ii}$\lambda$1808 and 1526, imposing the
presence of the main component with a fixed Doppler parameter. We
then use the resulting decomposition to fit the Fe~{\sc ii} lines.
Results are shown in Fig~\ref{dopplerb} for $b$~=~10 and 20 km/s. It
is apparent that components narrower than $b$~$\sim$~20~km/s are not
favored and $b$~=~10~km/s is definitely rejected. Therefore, we are
confident that imposing $b$~$\sim$~20~km/s for the main component in
the DLA will give us a good estimate of column densities.

For some of the species, several lines with very different
oscillator strengths (either in doublets or multiplets) are present
so that we can derive robust estimates of the column densities. This
is the case for Si~{\sc ii}, Si~{\sc ii}$^{*}$, and Fe~{\sc ii}.
From this, we derive metallicities relative to solar,
[Si/H]~=~$-$0.79 and [Fe/H]~=~$-$1.87. We also derive an upper limit
on $N$(Fe~{\sc iii}).

We also fit the high-ionization species. We did not try to tie the
components in different profiles because this high-ionization phase
could be highly perturbed. In Table~\ref{hiion}, except for the
first two components of the C~{\sc iv} absorption profiles, all
reported column densities are upper limits because the profiles are
strongly saturated. It can be seen however that the decompositions
in sub-components are fairly consistent between the different
species.

We detect absorptions from O~{\sc i}$^{**}$ and Si~{\sc ii}$^{*}$
which are rarely detected in DLAs (see Noterdaeme et al. 2015,
Neeleman et al. 2015) and are  more commonly  seen  in  DLAs
associated  to  GRB  afterglows (Vreeswijk et al.  2004;  Chen et
al.  2005;  Fynbo et al.  2006). Absorption from the fine structure
state of Si~{\sc ii} will be used to constrain the density of the
absorbing cloud (see below).

\newcommand{\avg}[1]{\left< #1 \right>} 

The high measured depletion of iron relative to silicon (i.e.
[Si/Fe]~=~+1.08) in this DLA suggests the presence of dust.
Consequently, extinction due to dust might be significant along this
line of sight. Indeed, the median g~$-$~r color for a sample of 697
non-BAL quasars with redshift within $\Delta$$z$~=~$\pm$0.02 around
$z_{\rm em}$~=~3.1910 is $\avg{g-r}$~=~0.30 when $g-r$~=~1.1 for QSO
J0823+0529. The reddening for this line of sight is E(B-V) = 0.09,
measured with an SMC reddening law template, which places it among
the most reddened of the sight lines in the Finley et al. (2013)
statistical sample. We will take this reddening into account in the
following while discussing the properties of the quasar.

\subsection{Physical conditions in the DLA gas}

In this section we use the photo-ionization code  {\sc cloudy} to
constrain the ionization state of the absorber and its distance to
the quasar central engine. Observed ionic ratios of Si~{\sc ii},
Si~{\sc iv}, Al~{\sc iii}, and Ar~{\sc i} are used to constrain the
plane parallel models constructed for a range of ionization
parameters. The calculations were stopped when a neutral hydrogen
column density of log~$N$(H~{\sc i})~=~21.70 is reached. Solar
relative abundances are assumed and the metallicity is taken to be
Z~=~0.16\,Z$_{\odot}$ from the Silicon abundance of the DLA derived
in Section~3.1.

\begin{table}
\centering \caption{Column densities of low-ionization species. The
quoted redshift is the redshift of the strongest component in the
low ion species. } \setlength{\tabcolsep}{8.0pt}
\renewcommand{\arraystretch}{1.2}
\begin{tabular}{c c  c c}
\hline

Redshift  &  Ion    &  log(N)$^a$           &  log(N) \\
$ $  &  $ $    &  $\rm Observed$    &  $\rm Model$ \\
\hline

3.190974 & Si~{\sc ii}           &  16.42$\pm$0.10   & 16.53   \\
  & Si~{\sc ii}$^{*}$     &  15.49$\pm$0.30   & 15.49   \\
  & Fe~{\sc ii}           &  15.33$\pm$0.20   & 16.50   \\
 & Fe~{\sc iii}          &  $\le$~15.00      & 15.09   \\
  & Ar~{\sc i}            &  14.97$\pm$0.20   & 15.04   \\
  & Al~{\sc iii}          &  14.80$\pm$0.10   & 14.64   \\
  & O~{\sc i}$^{**}$      &  16.45$\pm$0.45   & 15.70   \\
  & O~{\sc i}             &  17.08$\pm$0.50   & 17.62   \\

\hline \multicolumn{4}{l}{$^{a}$The Doppler parameter is fixed at
20~km~s$^{-1}$
(see Fig. 3).}\\
\end{tabular}
 \label{lowion}
 \renewcommand{\footnoterule}{}
\end{table}

\begin{table}
\centering \caption{Column densities of high-ionization species.}
\setlength{\tabcolsep}{2.0pt}
\renewcommand{\arraystretch}{1.3}
\begin{tabular}{c c c c c}
\hline
Redshift  &  Ion    & b  &  log(N)    &  log(N) \\
$ $  &  $ $    & [km~s$^{-1}$]  &  $\rm Observed$    &  $\rm Model$$^a$ \\
\hline
3.185037 & C~{\sc iv} & 23        &  13.60$\pm$0.10 & $ $   \\
3.186553 & C~{\sc iv} & 32        &  13.90$\pm$0.10 & $ $   \\
3.189333 & C~{\sc iv} & 86$\pm$10 &  $\ge$15.40$\pm$0.10 & $ $   \\
3.190953 & C~{\sc iv} & 38$\pm$14 &  $\ge$14.65$\pm$0.30 & $ $   \\
3.192634 & C~{\sc iv} & 39$\pm$6  &  $\ge$15.40$\pm$0.50 & $ $   \\
Total(C~{\sc iv})  &  &           &  $\ge$15.75          & 18.43 \\
\hline
3.189333 & Si~{\sc iv} & 56$\pm$10 &  $\ge$14.30$\pm$0.10 & $ $   \\
3.190953 & Si~{\sc iv} & 25$\pm$14 &  $\ge$14.60$\pm$0.60 & $ $   \\
3.192634 & Si~{\sc iv} & 25$\pm$6  &  $\ge$15.72$\pm$0.90 & $ $   \\
Total(Si~{\sc iv})  &  &           &  $\ge$15.75          & 16.14 \\
\hline
3.189821 & N~{\sc v} & 99$\pm$39 &  $\ge$14.80$\pm$0.30 & $ $   \\
3.191157 & N~{\sc v} & 34$\pm$29 &  $\ge$14.89$\pm$0.20 & $ $   \\
3.192807 & N~{\sc v} & 74$\pm$19  &  $\ge$14.87$\pm$0.10 & $ $   \\
Total(N~{\sc v})  &  &           &  $\ge$15.33          & 16.55 \\
\hline
3.189788 & P~{\sc v} & 136 &  $\ge$14.20$\pm$0.20 & $ $   \\
3.191109 & P~{\sc v} & 97  &  $\ge$13.95$\pm$0.30 & $ $   \\
3.192757 & P~{\sc v} & 13  &  $\ge$13.80$\pm$0.60 & $ $   \\
Total(P~{\sc v})  &  &           &  $\ge$14.50          & 15.30 \\
\hline

\multicolumn{5}{c}{$^{a}$The Model values are for the ionization
parameter logU~=~$-$0.3.} \\
\end{tabular}
 \label{hiion}
\end{table}

\begin{figure}
\centering
\begin{tabular}{c}
\includegraphics[bb=60 374 248 652,clip=,width=0.95\hsize]{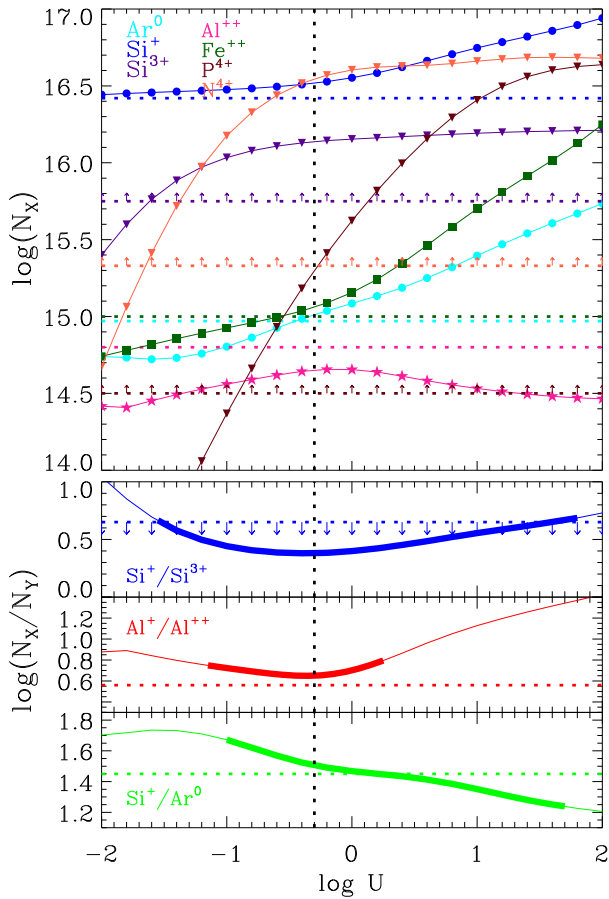}
\end{tabular}
\caption{Results of {\sc cloudy} models of the DLA system. Bottom
panel gives resulting ionic ratios and top panel shows colum
densities. In both panels measurements are shown by dashed
horizontal lines and upper limits are indicated with arrows. The
thick lines in the bottom panel show the range of ionization
parameters constrained by each ionic ratio. }
 \label{Model}
\end{figure}

\begin{table}
\centering \caption{Hydrogen density ($n_{\rm H}$) in the cloud, its
distance to the central AGN ($r$$_{0}$), size of the DLA ($l_{\rm
HI}$) and transverse size of the C~{\sc iv} phase for different
values of the ionization parameter (U). }
\setlength{\tabcolsep}{7.0pt}
\renewcommand{\arraystretch}{1.2}
\begin{tabular}{c c c c c}
\hline

log~U  &  $n$$_{\rm H}$    &  $r$$_{0}$ &  $l$(H~{\sc i}) & $l$(C~{\sc iv}) \\
$ $  &  [cm$^{-3}$]   &  [pc]      &  [pc]  & [pc]\\
\hline

$-$1.1 & 710 &  579 & 2.3  & 3.0 \\
$-$0.7 & 500 &  435 & 3.2  & 5.8 \\
$-$0.3 & 355 &  326 & 4.6  & 14.6 \\
  +0.0 & 250 &  275 & 6.5  & 33.7\\
  +0.3 & 180 &  229 & 9.1  & 80.0\\

\hline
\end{tabular}
 \label{DLAsize}
 \renewcommand{\footnoterule}{}
\end{table}

The ionizing spectrum incident on the cloud is taken as the
combination of the standard AGN spectrum of Mathews \& Ferland
(1987), Haardt-Madau extragalactic spectrum (Haardt \& Madau 2005,
HM05) and the CMB radiation both at $z$~=~3.1910. Fig.~\ref{Model}
summarizes the results of these calculations. Bottom panel gives
resulting ionic ratios and top panel shows column densities. In both
panels measurements and upper limits are indicated by dashed lines.
For Al~{\sc ii}, we scale the Si~{\sc ii} column density assuming
solar metallicity ratios.

The column density ratio log$N$(Al~{\sc ii})/$N$(Al~{\sc iii})
yields ionization parameter ranging from $-$1.1 to 0.3. Both
log\,$N$(Si\,{\sc ii})/$N$(Ar\,{\sc i}) ratio and the limit on
log\,$N$(Si\,{\sc ii})/$N$(Si\,{\sc iv}) are consistent with this
range of ionization parameters. Our preferred value is
log~$U$~=~$-0.3$ and we indicate the column densities for this model
in Tables~1 and 2. Note that we detect a trough at the position of
P\,{\sc v}$\lambda$1117 with an absorption profile which is
consistent with that of other high-ionization species (see Fig.~2).
The second weaker line of the doublet, P\,{\sc v}$\lambda$1128 is
affected by noise. The fit to these two lines gives a column density
which is consistent with the results of the preferred model. The
same is true for N~{\sc v}. This strongly supports the fact that the
cloud is highly ionized.

To determine the hydrogen density of this cloud (i.e. $n_{\rm H}$),
a series of {\sc cloudy} models with fixed ionization parameter
(varying from log~U~=~$-$1.1 to 0.3) and varying $n_{\rm H}$ were
constructed. Note that in this series of models, the metallicity and
incident radiation are the same as those considered above. By
increasing the hydrogen density we are indeed trying to
collisionally populate the excited states of the Si~{\sc ii} ground
state to explain the observed Si~{\sc ii}$^*$ column density.

Knowing the ionization parameter and the density, we can derive the
distance of the cloud to the quasar by estimating the number of
ionizing photons emitted by the quasar. Since the quasar is
reddened, we first estimate the flux at 20370\AA~ (corresponding  to
H$\beta$ at the redshift of the quasar) by extrapolating with a
power-law the continuum observed at 6125 and 8165~\AA. Following
Srianand \& Petitjean (2000), we then assume that the de-reddened
quasar continuum is a power law
($f_{\lambda}$~$\sim$~$\lambda^{\alpha_{\lambda}}$; with
$\alpha_{\lambda}$~=~$-$1.5) and we consider that the flux at
$\sim$~20370~$\textup{\AA}$ is not affected by the reddening. We
thus estimate the flux at the Lyman limit in the rest frame of the
absorber to be $f_{912{\tiny\textup{\AA}}}$~=~3.40$\times$10$^{-17}$
erg~s$^{-1}$cm$^{-2}$$\textup{\AA}^{-1}$. This flux corresponds to a
luminosity of $L_{912{\tiny\textup{\AA}}}$~=~3.25$\times$10$^{42}$
erg~s$^{-1}$$\textup{\AA}$$^{-1}$ at the Lyman limit. We can now
estimate the rate at which hydrogen ionizing photons are impinging
upon the face of the cloud by integrating $L_{\nu}$/$h\nu$ over the
energy range 1 to 20 Ryd. If we assume a flat spectrum for the
quasar over this energy range, we get $Q$~=~6.78$\times$10$^{55}$
photons per second. From the definition of the ionization parameter
U,

\begin{equation}
U = \frac{Q}{4 \pi r_{0}^{2} n_{\rm H} c}
\end{equation}

\par\noindent
one can see that for given values of $n_{\rm H}$, U, and $Q$, one
can uniquely estimate the distance, $r_{0}$, from the quasar to the
absorbing cloud ($c$ is the speed of light).

We can then derive the size of the cloud along the line of sight.
For the neutral part, the size along the line of sight is
$l$~$\sim$~$N_{\rm HI}$/$n_{\rm H}$. Results are summarized in
Table~\ref{DLAsize} which gives the hydrogen density, distance to
the quasar and size of the neutral phase of the cloud for different
values of log~$U$. The longitudinal size range from 2.3 to 9.1~pc
for a distance of, respectively, 579 to 229~pc from the quasar and a
hydrogen density of 710 to 180 cm$^{-3}$.
Using the structure given by the model and assuming that the cloud
is spherical we can derive the transverse size of the C~{\sc iv}
phase. It is given in the last column of Table~\ref{DLAsize} and
range from 3 to 80~pc. These estimates for the transverse size of
the high-ionization zone are only rough estimates because our model
is very simple (only one density) and we assume spherical geometry.

\begin{figure}
\centering
\includegraphics[bb=115 357 458 613,clip=,width=0.95\hsize]{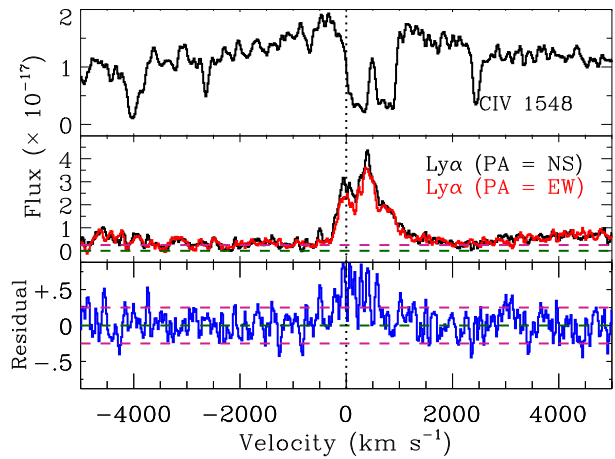}
\caption{Top panel shows on a velocity scale the quasar C~{\sc iv}
emission and C~{\sc iv} absorptions from the DLA. The Lyman-$\alpha$
regions for the two PAs (PA~=~NS in black, PA~=~EW in red) are
overplotted in the middle panel. The DLA trough never reaches the
zero level. Bottom panel shows the subtraction of the two spectra
(note the change in the y-scale). The residual is consistent in the
two spectra. Note that the y-axis is in
erg~s$^{-1}$~cm$^{-2}~$\textup{\AA}$^{-1}$. }
 \label{Lya_residual_2PA}
\end{figure}

\subsection{Residual flux in the bottom of the DLA trough}

We observe residual flux in the bottom of the DLA trough extending
in velocity well beyond the narrow emission line. This can be seen
in Fig.~\ref{Lya_residual_2PA} where we overplot zooms of the
Lyman-$\alpha$ regions observed along the two PAs. It is apparent
that the flux is never at zero in the trough. We checked in the
Lyman-$\alpha$ forest that the bottom of saturated lines have on an
average zero flux. The residual flux is consistent in these two
spectra as is demonstrated in the bottom panel of the figure which
gives the difference between the two spectra.

This means that the DLA cloud does not cover the background source
completely. Since the source of the quasar continuum is much smaller
than the broad line region (e.g. Hainline et al. 2013), this excess
is due either to the cloud not covering the BLR completely or to a
second continuum source.

We investigate here if this excess can be due to an additional
continuum source. If this excess is due to the continuum of the host
galaxy then we would expect some residual at the bottom of metal
lines. Indeed C~{\sc ii}$\lambda$1334 is saturated and seems to show
some weak residual (see Fig.~2). We thus have scaled the LBG mean
continuum as given by Kornei et al. (2010) to the bottom of C~{\sc
ii}$\lambda$1334. We find that this could explain 20 to 50\% of the
DLA residual. The corresponding magnitude of the galaxy would be
23.22 which would be very bright. In any case this possibility
cannot explain all the residual.

\begin{figure}
\centering
\begin{tabular}{c}
\includegraphics[bb=117 385 458 652,clip=,width=0.95\hsize]{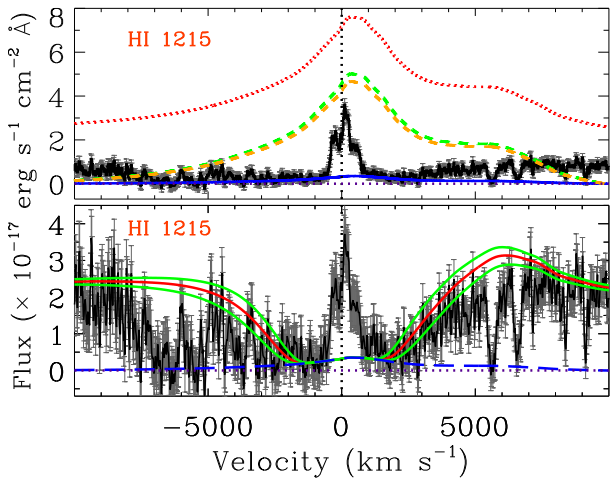}
\end{tabular}
\caption{{\sl Upper panel}: The residual in the bottom of the DLA
trough seen over the velocity range [$-$3000, +3000]~km~s$^{-1}$ can
be explained if $\sim$10\% of the broad line region is not covered
(solid blue line). The de-reddened QSO spectrum (dotted red curve)
recovered from PCA analysis of the red part of the spectrum is  used
to estimate the quasar emission lines (dashed green line) that are
then scaled to derive the covering factor of the cloud. {\sl Lower
panel}: The red curve is the DLA fit conducted on the de-reddened
observed spectrum. Green curves indicate the $\pm$1$\sigma$ limits.}
 \label{simulateHI}
\end{figure}

It is probable that part of the residual flux in the bottom of the
DLA trough is due to the fact that part of the Lyman-$\alpha$
emission from the BLR is not covered by the DLA. The size of the
broad line region (BLR) can be inferred from time delay measurements
between variations in the continuum and in the broad lines. Recent
investigations of low-redshift AGNs show a tight relation between
this size and the luminosity of the AGN, $R=A\times (L/10^{43})^B$,
where $R$ is the radius of the BLR, $A$ is a typical distance in
light-days and $L$ is the luminosity either in an emission line
(H$\beta$ or C~{\sc iv}) or in the continuum. The index is found to
have a value close to B$\sim$0.6-0.7 when the typical distance A is
in the range 20-80 light-days for local AGNs (Wu et al. 2004; Kaspi
et al. 2005). More recently Bentz et al. (2009) find log $(R_{\rm
BLR})=K+\alpha {\rm log}(\lambda L_{\lambda}$(5100\AA)) with
$\alpha$~=~0.519$^{+0.063}_{-0.066}$ and
K~=~$-$21.3$^{+2.9}_{-2.8}$. The slope suggests that brighter AGNs
have to a first approximation the same structure as fainter AGNs
with only larger dimensions. Therefore, extending the Kaspi et al.
(2005) relation to higher luminosities yields a typical radius of
the order of 1 pc for the BLR of bright high-z quasars. In the
present case, luminosity is $L_{\rm CIV}~\sim~9~\times~10^{44}$
erg~s$^{-1}$ which gives a size of the BLR of $\le$~1.1~pc. We have
estimated in the previous Section that the longitudinal size of the
DLA is of order 2-9~pc. This is larger than the size of the BLR. It
therefore would mean either that the cloud is much smaller in the
transverse direction, corresponding possibly to a filamentary
structure or that some holes are present in the cloud or that the
cloud is not centered on the quasar.

\subsection{QSO BLR Lyman-$\alpha$ emission and covering factor}

To determine the H~{\sc i} column density accurately, one needs to
reconstruct the shape of the quasar spectrum at the position of the
DLA profile.  We can apply principal component (PCA) reconstruction
of the quasar flux using the red part of the spectrum to estimate
the shape of the Lyman-$\alpha$$-$N~{\sc v} emission (P\^aris et al.
2011, 2013). To do so, we first subtract the residual flux seen at
the bottom of the DLA (i.e. residuals from BEL and NEL regions) so
that we get zero flux in the DLA trough. We then de-redden this
spectrum and add again the residuals subtracted above. We now have
the complete de-reddened observed spectrum. Applying the Principle
Component Analysis (PCA) method on this spectrum, we derive the PCA
spectrum shown as the dotted red curve in the upper panel of
Fig.~\ref{simulateHI}.
Then, we subtract the continuum from the PCA spectrum (green dashed
curve) and scale it so that it is consistent with the residual flux
seen at the bottom of the DLA trough (solid blue curve).
This residual flux is only $\sim$7~\% of that of the Lyman-$\alpha$
broad emission, indicating that $\sim$93~\% of the Lyman-$\alpha$
BLR is covered by the cloud.

To obtain the flux seen by the DLA, we subtract from the dashed
green curve the residual flux seen at the bottom of the DLA (i.e.
the solid blue curve) and re-add the continuum. We then fit the DLA
to obtain the solid red curve in the lower panel. We also checked
that we get the same result if, before fitting the DLA, we first
re-redden the PCA continuum. As can be seen in
Fig.~\ref{simulateHI}, the fit is slightly high near the N~{\sc v}
emission line although within errors. This is possibly because the
N~{\sc v} emission line in this quasar may be weaker than what is
predicted by the PCA method. It should be reminded that the PCA
reconstruction is an estimate of the quasar spectrum which stays
close to the median spectrum in the overall quasar population (see
P\^aris et al. 2011). The result (log~$N$(H~{\sc i})~=~21.7$\pm$0.1)
is consistent with what was derived previously.

\begin{figure}
\centering
\begin{tabular}{c}
\includegraphics[bb=110 390 458 577,clip=,width=0.95\hsize]{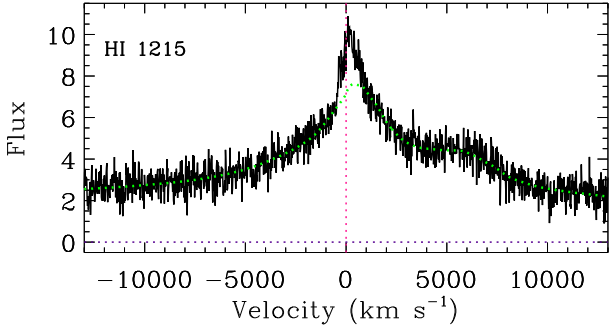}
\end{tabular}
\caption{Reconstructed quasar spectrum as it would be observed if no
absorption were present. The original PCA spectrum is overplotted as
a dotted green curve. The y-axis is in
erg~s$^{-1}$~cm$^{-2}$\textup{\AA}$^{-1}$. }
 \label{simulateQSO}
\end{figure}

\begin{figure}
\centering
\begin{tabular}{c}
\includegraphics[bb=177 354 407 721,clip=,width=0.95\hsize]{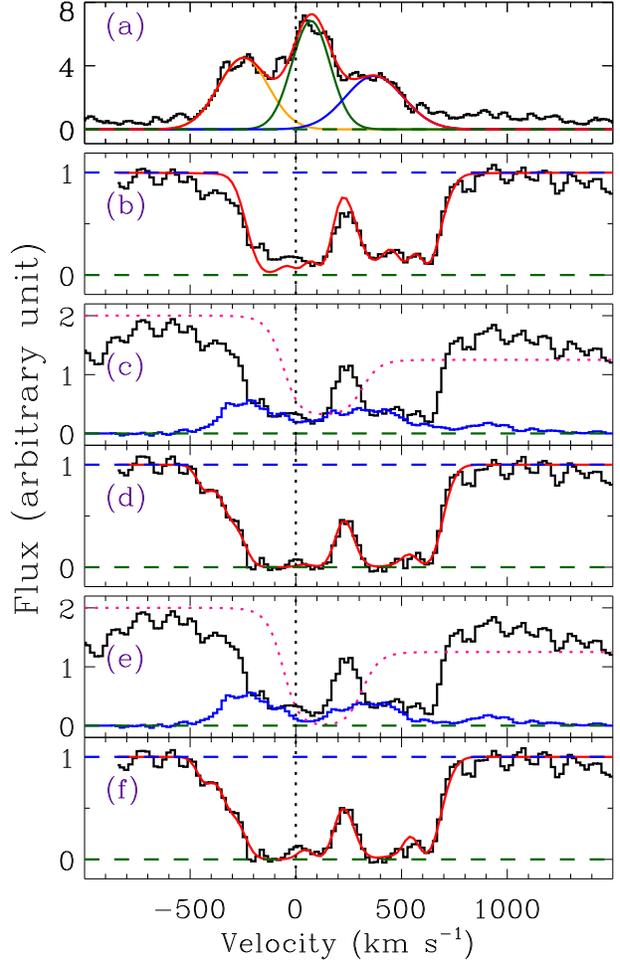}
\end{tabular}
\caption{{\sl Panel a}: Lyman-$\alpha$ emission line profile
detected in the trough of the DLA. Three Gaussian functions are used
to fit the profile. {\sl Panel b}: C~{\sc iv} absorption profiles
(the two absorption lines C~{\sc iv}$\lambda$1548 and C~{\sc
iv}$\lambda$1550 are seen). It is apparent from the flat bottom of
the lines and similar apparent optical depths that the covering
factor of the C~{\sc iv} gas is not unity. This is demonstrated by
the model fit overplotted as a red line. Note that a Fe~{\sc
ii}$\lambda$2586 component at $z_{\rm abs}$~=~1.5140 is blended in
the red wing of the C~{\sc iv}$\lambda$1550 profile. {\sl Panel c}:
The residual at the bottom of the C~{\sc iv} lines is fitted by
scaling the Lyman-$\alpha$ emission profile. The red dotted line
indicates (on an arbitrary scale) the scaling factor (see Eq.~(4)).
The effective values of the C~{\sc iv}/Lyman-$\alpha$ ratio are
fitted to be 0.29, 0.045 and 0.18 for the three emission components
respectively. The C~{\sc iv}/Lyman-$\alpha$ ratio must be much
smaller in the second emission component. {\sl Panel d}: Fit of the
C~{\sc iv} doublet once the residuals (including both C~{\sc iv}
emission lines) are removed. {\sl Panel e}: Same as Panel (c) but
the central Lyman-$\alpha$ component is supposed to have no C~{\sc
iv} associated. {\sl Panel f}:  Fit of the C~{\sc iv} doublet once
the residuals from panel (e) are removed (no C~{\sc iv} is
associated with the central Lyman-$\alpha$ component). }
 \label{CIV_NEL_II}
\end{figure}

Figure~\ref{simulateQSO} shows the quasar spectrum as it would be
observed if no DLA were present. In this figure, we have added the
Lyman-$\alpha$ narrow emission component seen in the DLA trough to
the PCA spectrum. The final spectrum is degraded to the
SNR~$\sim$~20 by adding Gaussian noise to it.
It is apparent from this spectrum that the emission from the narrow
line region is quite strong in this quasar.

\section{The quasar narrow line region}

We call here the narrow line region (NLR), the region of the quasar
host galaxy that is located within the PSF of the observations. This
corresponds to about 1~arcsec or 7.1~kpc at the redshift of the
quasar (or a distance of 3.55~kpc on both sides of the quasar). The
emission seen beyond this will be called extended emission. We have
shown that the DLA is of much smaller dimension so that most of the
Lyman-$\alpha$ NLR is not covered and is detected as Lyman-$\alpha$
emission in the bottom of the DLA trough. We can extract this
emission which is shown in the top panel of Fig.~\ref{CIV_NEL_II}.
It can be seen that the emission is spread over more than
1200~km~s$^{-1}$ with FWHM~$\sim$~900~km~s$^{-1}$.

\subsection{CIV partial coverage}

The large rest equivalent width of the C~{\sc iv} and Si~{\sc iv}
absorption doublets and the flat-bottomed structure of their
profiles suggest that these lines are saturated (see
Fig.~\ref{CIV_NEL_II}). However, the flux at the bottom of the
C~{\sc iv} doublet absorption lines apparently does not reach the
zero flux level, indicating that the absorbing cloud is only
partially covering the background emission-line region.

When partial coverage occurs, the residual intensity seen at the
bottom of absorption lines can be written as at each wavelength:
\begin{equation}
I(\lambda)~=~I_{0}(\lambda)(1~-~C_{\rm f})~+~C_{\rm
f}~I_{0}(\lambda)~exp~[-\tau(\lambda)]
\end{equation}

\par\noindent
where $I_{0}(\lambda)$ is the incident (unabsorbed) intensity,
$\tau(\lambda)$ is the optical depth of the cloud at the considered
wavelength, and $C_{\rm f}$ is the fraction of the background
emitting region that is covered by the absorbing cloud (i.e. the
covering factor). In the case of doublets, and assuming the covering
factor is the same for each component of the doublet,  we can write:

\begin{equation}
C_{\rm f}~=~\frac{1~+~R_{2}^{2}~-~2R_{2}}{1~+~R_{1}~-~2R_{2}}
\end{equation}

\par\noindent
where $R_{1}$ and $R_{2}$ are the normalized residual intensities in
the two absorption lines of the doublet (Petitjean \& Srianand 1999;
Srianand \& Shankaranarayanan 1999).

\begin{figure*}
\centering
\begin{tabular}{c}
\includegraphics[bb=52 401 557 537,clip=,width=0.95\hsize]{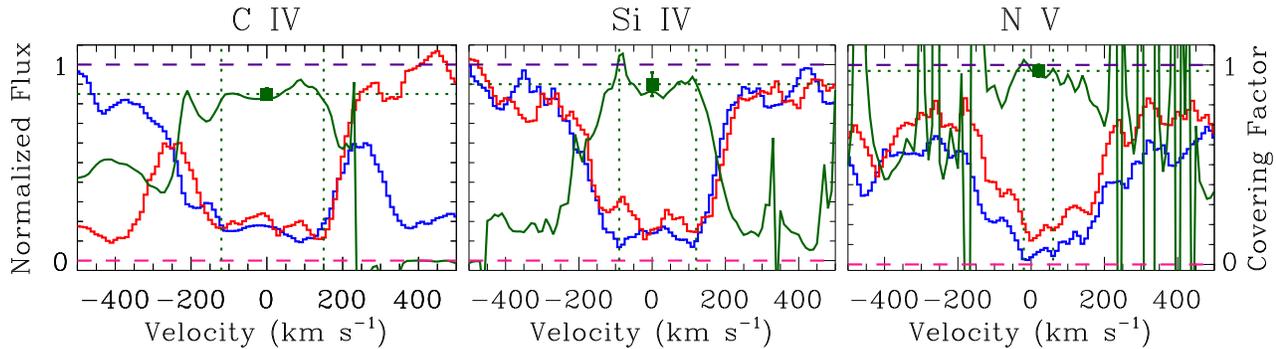}
\end{tabular}
\caption{Covering factor estimations for C~{\sc iv}, Si~{\sc iv},
and N~{\sc v} doublets following Eq.(3). Vertical dashed lines
indicate the windows in which it is calculated and horizontal dashed
lines (and filled squares) give the mean over these windows. The red
and blue histograms are the profiles of the weaker and stronger
transitions of each doublet, respectively. }
 \label{CIV_Cf}
\end{figure*}

In Fig.~\ref{CIV_Cf}, we show the covering factor of the C~{\sc iv},
Si~{\sc iv}, and N~{\sc v} doublets. Here, the red and blue
histograms are the profiles of the weaker and stronger transitions
of each doublet, respectively. The green solid line is the covering
factor at each point of the profile calculated from Eq.(3). The
green vertical dotted lines mark the regions used to calculate the
mean values of $C_{\rm f}$, avoiding the wings of the profiles. The
green horizontal dotted lines along with the green filled squares
show these mean values. Error bars are calculated as the standard
deviation of $C_{\rm f}$ in different velocity bins. The measured
values of $C_{\rm f}$ for C~{\sc iv} and Si~{\sc iv} are $\sim$~0.85
and 0.90 respectively suggesting that the size of the corresponding
BEL region may be similar for the two high-ionization species. Note
also that N~{\sc v} on the contrary seems fully covered.

The above numbers are however only indicative because the resolution
of our spectrum is not high ($R\sim 4000$). In addition there is a
possible FeII$\lambda$2586 line at $z_{\rm abs}~=~1.5140$ blended
with the red wing of C~{\sc iv}$\lambda$1550 (cyan curve in the
C~{\sc iv}$\lambda$1550 panel in Fig~\ref{j08231D}). In
Fig.~\ref{CIV_NEL_II}, we demonstrate the effect of partial coverage
on the observed C~{\sc iv} doublet profiles taking into account the
resolution of the spectrum and the presence of the FeII line. The
contribution of the Fe~{\sc ii}$\lambda$2586 absorption to this fit
is robustly determined by fitting together Fe~{\sc
ii}$\lambda$$\lambda$2344,2600 in the same system. In panel (b) of
Fig.~\ref{CIV_NEL_II} the red curve is the fit conducted with
VPFIT\footnote{ \url{http://www.ast.cam.ac.uk/~rfc/vpfit.html}} on
the original data. Here, we can see that there is no way to fit most
of the profile without invoking partial coverage.

\subsection{Reconstructing C~{\sc iv} narrow emission line (NEL) profile}

As mentioned earlier, the flux at the bottom of the C~{\sc iv}
doublet absorption lines does not reach the zero level. This
residual flux could be due to the partial coverage of either the
C~{\sc iv} BLR or NLR. However, the radius of the C~{\sc iv} phase
associated with the DLA is larger and sometimes much larger than the
size of the BLR (see Section 3.2) which strongly suggests that the
residual flux is due to the C~{\sc iv} NLR.

We thus tried to reproduce the residual by C~{\sc iv} emission,
including {\sc both} lines of the doublet. Doing this we would like
to ask the question whether we can associate the emission with the
Lyman-$\alpha$ emission. Indeed, this could bring important clues on
the origin of the Lyman-$\alpha$ emission. In case the
Lyman-$\alpha$ emission {\sl cannot} be associated with C~{\sc iv}
emission then this would strongly suggest that the Lyman-$\alpha$
emission corresponds to scattered light after radiative transfer.

We thus first try to simply scale the Lyman-$\alpha$ profile. This
is bound to fail as it is apparent that the residuals do not follow
the shape of the Lyman-$\alpha$ emission. We therefore decomposed
the Lyman-$\alpha$ profile in three Gaussian functions (see top
panel of Fig.~\ref{CIV_NEL_II}) as suggested by the profile itself.

This allows us to scale different parts of the Lyman-$\alpha$
profile differently. For each pixel we define the ratio $R$~=~C~{\sc
iv}/Lyman-$\alpha$ by combining the three Gaussian functions using
the following equation:

\begin{equation}
R~=~\frac{R_{1}\times G_{1}~+~R_{2}\times G_{2}~+~R_{3}\times
G_{3}}{G_{1}~+~G_{2}~+~G_{3}}
 \label{eq_CIV_NEL}
\end{equation}

\par\noindent
where  $R_{1}$, $R_{2}$, and $R_{3}$ are the weights for each
component directly related to the C~{\sc iv}/Lyman-$\alpha$ ratio in
this component. By assigning different values to $R_{1}$, $R_{2}$,
and $R_{3}$, we can scale each Gaussian individually. For instance,
if $R_{1}$~=~$R_{2}$~=~$R_{3}$~=~1.0, this results in the red
profile shown in panel (a) of Fig.~\ref{CIV_NEL_II}. which is simply
the combination of the three Gaussian functions. One can now use the
factor $R$~=~C~{\sc iv}/Lyman-$\alpha$ defined for each wavelength
to properly scale the Lyman-$\alpha$ NEL profile. We thus have to
adjust the $R_{1}$, $R_{2}$, and $R_{3}$ parameters until the C~{\sc
iv} emission is consistent with the residual seen at the bottom of
the C~{\sc iv} doublet absorption lines.

We find that the residual can be reproduced (see panel (c) of
Fig.~\ref{CIV_NEL_II}) with $R_{1}$~=~0.29, $R_{2}$~=~0.045, and
$R_{3}$~=~0.18. The variation of $R$ through the profile is given as
a dashed pink line in panels (c) and (e) of Fig.~\ref{CIV_NEL_II}.
The corresponding fit of C~{\sc iv} after removing the residuals is
given in panels (d) and (f).

Since the weight of the second emission component is much smaller
than the two other ones, we ask the question whether it would be
possible that the second component has no C~{\sc iv} associated. To
test this, we impose $R_{2}$~=~0.001. The result of the fit is given
in panels (e) and (f). It is apparent that we can find a solution
with no C~{\sc iv} associated with the second component.

We thus conclude from all this that (i) the C~{\sc iv} emission is
strongest around $v$~$\sim$~$-$200~km~s$^{-1}$; this position could
indicate the redshift of the quasar, implying that most of the
Lyman-$\alpha$ emission and the DLA are redshifted; (ii) at least
part of the NLR Lyman-$\alpha$ emission has no C~{\sc iv} emission
associated (predominantly around $v=+100$~km~s$^{-1}$) which means
that the Lyman-$\alpha$ emission in this component is due to
scattered light or that the emitted gas is located within a distance
from the quasar smaller than the transverse size of the C~{\sc iv}
phase associated with the cloud so that the corresponding C~{\sc iv}
emission is absorbed by the high-ionization phase of the DLA cloud.

\section{Extended Lyman-$\alpha$ emission}

\begin{figure*}
\centering
\begin{tabular}{c}
\includegraphics[bb=68 386 546 663,clip=,width=0.80\hsize]{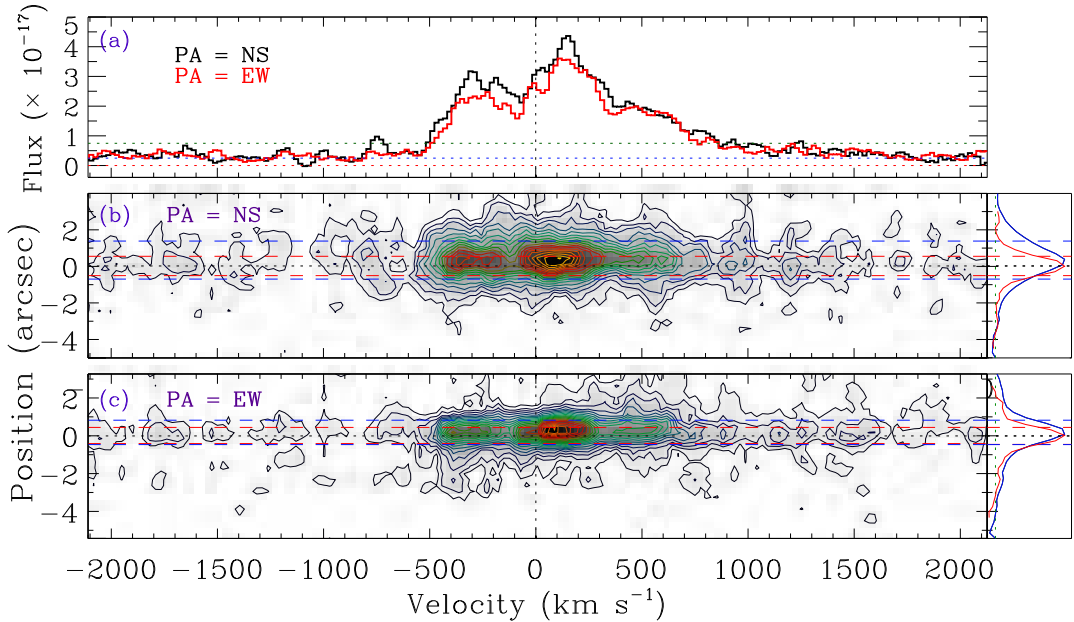}
\end{tabular}
\caption{{\sl Panel~(a)}: 1D spectrum showing the Lyman-$\alpha$
emission in the bottom of the DLA trough. Black and red histograms
are for PA~=~NS and EW, respectively. Red, blue and green horizontal
dotted lines mark the zero flux level, 1~$\sigma$, and 3~$\sigma$
uncertainty on the flux, respectively. {\sl Panels~(b) and (c)}: 2D
spectra of the Lyman-$\alpha$ emission seen in the trough of the DLA
at $z_{\rm abs}$~=~3.1910 for PA~=~NS (panel b) and PA~=~EW (panel
c). The contours show the observed flux density with a separation
between two lines of 3.5~$\times$~10$^{-19}$
erg~s$^{-1}$~cm$^{-2}$~$\textup{\AA}^{-1}$. The outermost contour
corresponds to a flux density of 4.37~$\times$~10$^{-19}$
erg~s$^{-1}$~cm$^{-2}$~$\textup{\AA}^{-1}$. The black horizontal
dotted line shows the center of the continuum trace. The blue and
red dashed horizontal lines show the FWHM of the trace and extended
Lyman-$\alpha$ emission, respectively. Panels on the right show the
spatial emission profile integrated over the whole wavelength range
(blue curves) together with the continuum PSF (red lines) derived
from the red part of the order. Note that in {\sl Panel~(a)}, the
y-axis is in erg~s$^{-1}$~cm$^{-2}$~$\textup{\AA}^{-1}$. }
 \label{2D_both_PAs}
\end{figure*}

In the middle and bottom panels of Fig.~\ref{2D_both_PAs}  we show
the 2D spectra of the Lyman-$\alpha$ emission detected in the DLA
trough for the two PAs. It is apparent that the Lyman-$\alpha$
emission is extended and slightly displaced relative to the quasar
trace. To quantify this we integrate the whole Lyman-$\alpha$
profile in the spectral direction and compare the result to the
spatial PSF derived from the integration of the quasar spectrum over
the rest of the order beyond 6520~\AA. This is shown in the right
panels in Fig.~\ref{2D_both_PAs}. It is apparent that the emission
is extended well beyond the PSF for both PAs and mostly in one
direction implying that the total emission is shifted towards this
direction.

We investigate whether the extension of the emission varies with the
velocity position. For this, we split the velocity range over which
the emission is seen in several regions, following the profile, and
integrate the spatial emission profile over these regions. We then
fit the profile by a Gaussian function. The results show that along
PA~=~NS the spatial extent of the emission is larger than 5~arcsec
over about 2000~km~s$^{-1}$ and that the shift is about 0.2~arcsec
towards the North direction in the same region. For PA~=~EW,
extension is about the same but the shift is consistent with zero
meaning that the extended emission is more symmetric around the
trace. The Lyman-$\alpha$ emission is detected up to more than
25~kpc from the quasar and there is a strong excess emission along
PA~=~NS to the North.

To better visualize the extended emission, we will subtract the
emission associated with the central PSF e.g. the emission located
on the quasar trace. To do so, the 2D spectral order outside the
Lyman-$\alpha$ region of each PA is split into several chunks, and
in each chunk the counts are integrated along the spectral axis. A
Gaussian is then fitted on the profiles to get the central pixel
values of the spatial profiles. Finally, we fit a straight line on
these values to determine the center of the trace for each spectral
pixel.

\begin{figure*}
\centering
\begin{tabular}{c c}
\includegraphics[bb=114 396 458 667,clip=,width=0.46\hsize]{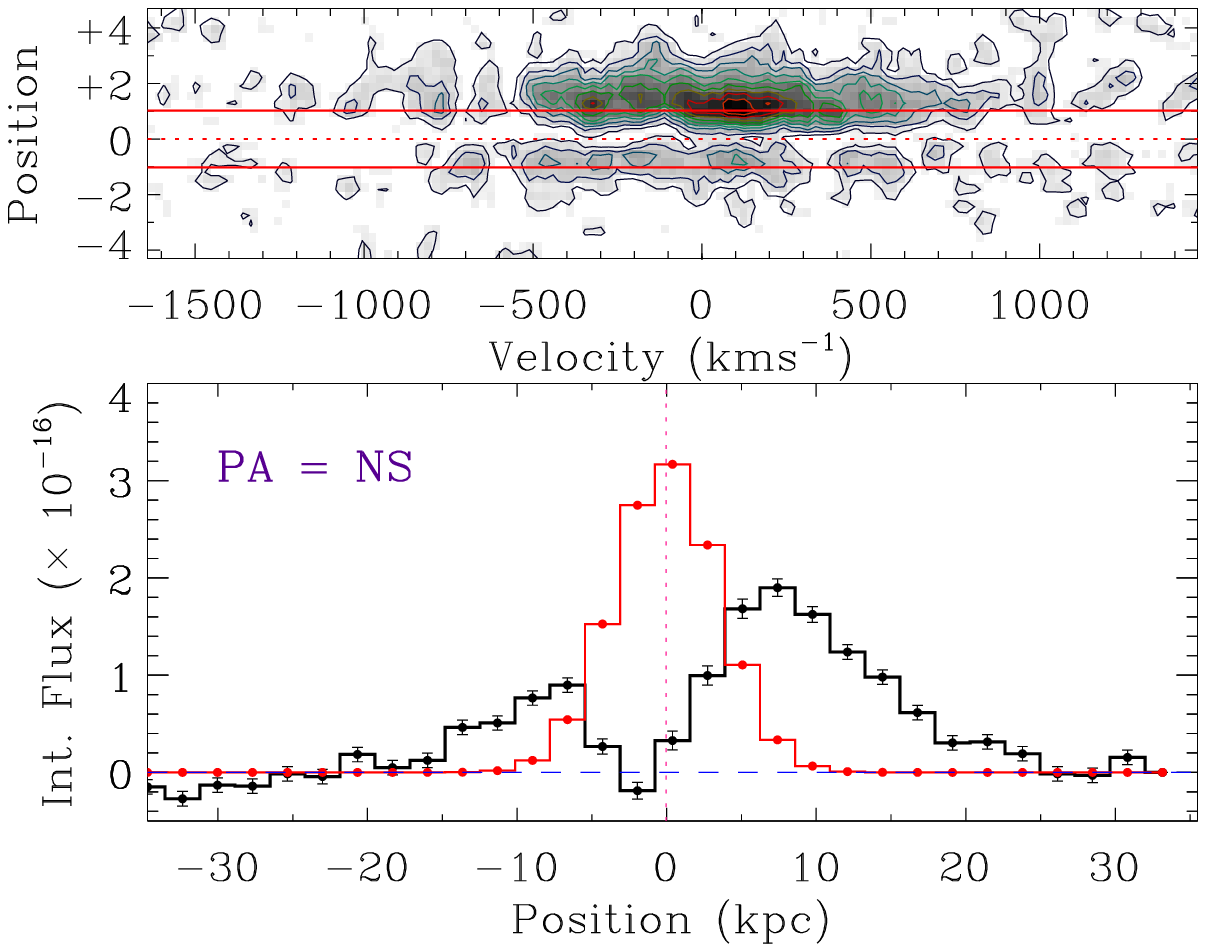}
\includegraphics[bb=114 396 459 667,clip=,width=0.46\hsize]{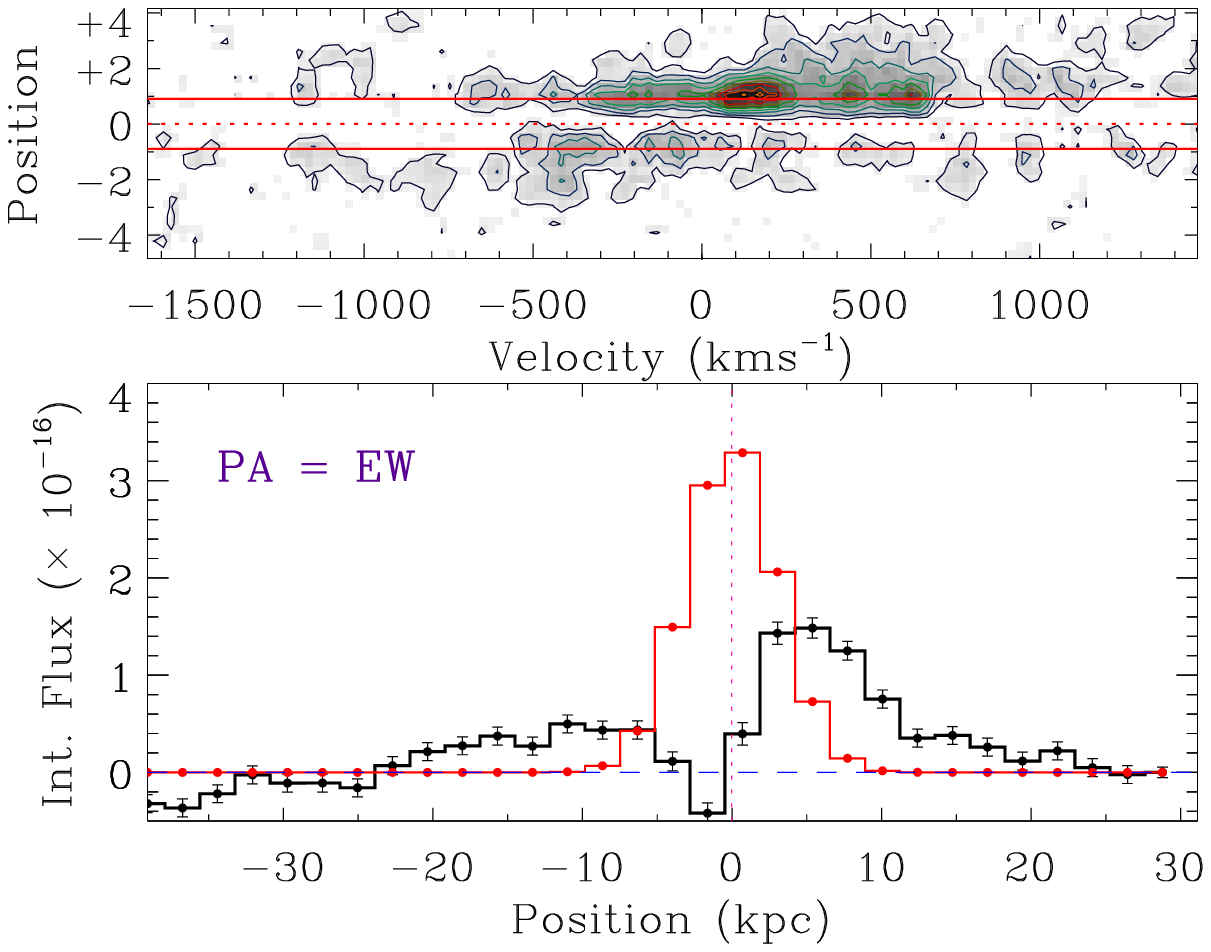}
\end{tabular}
\caption{{\sl Upper panels}: 2D spectra of the Lyman-$\alpha$
emission seen in the trough of the DLA (left PA~=~NS and right
PA~=~EW) after subtracting the emission in the PSF centered on the
quasar trace. The contour levels are the same as in
Fig~\ref{2D_both_PAs}. The red horizontal dotted line shows the
center of the trace and the red solid horizontal lines show the FWHM
of the trace. {\sl Bottom panels}: Emission profiles integrated over
the whole wavelength window of the Lyman-$\alpha$ emission in the
PSF centered on the quasar trace (red histograms) and the extended
emission (black histograms). Note that in the upper (resp. lower)
panels, the y-axis is in arcsec (resp.
erg~s$^{-1}$~cm$^{-1}$~arcsec$^{-2}$). }
 \label{Lya_vs_trace_extension}
\end{figure*}

Once we know the position of the trace exactly, we extract the
spatial profile in each wavelength pixel and we fit a Gaussian with
width equal to the continuum PSF width. We then subtract this
Gaussian from the profile. Results are given in
Fig~\ref{Lya_vs_trace_extension}. Top panels show the resulting 2D
spectrum of the extension. Bottom panels show the Lyman-$\alpha$
profiles integrated along the spectral direction: in red the
emission which was fitted on top of the trace and in black is the
extension.

\begin{figure}
\centering
\begin{tabular}{c}
\includegraphics[bb=119 376 458 721,clip=,width=0.95\hsize]{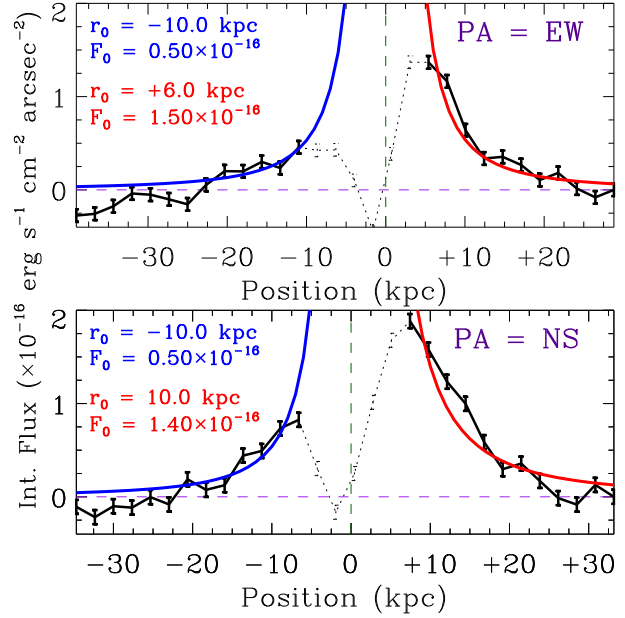}
\end{tabular}
\caption{Variations of the integrated Lyman-$\alpha$ flux with
distance from the AGN. The black lines with error bars are the
observations and the red and blue curves indicate a variation of the
flux with distance following a $r^{-2}$ relation. Note that in both
panels F$_{0}$ is in erg~s$^{-1}$~cm$^{-2}$~arcsec$^{-2}$. }
 \label{Distance}
\end{figure}

Along PA~=~NS, most of the extended emission is seen to the North up
to projected distances of 3.5~arcsec or $\sim$25~kpc from the
center. The striking observation here is that the emission is
extended over more than 1000~km~s$^{-1}$ at all distances from the
quasar. Along PA~=~EW, extension is more apparent in one direction
as well, to the West. However the velocity extension gets smaller
when the distance to the quasar increases. Fig~\ref{Distance} shows
the variation of the integrated flux with distance. In this Figure,
the red and blue curves indicate variations of flux with distance
following the relation $F$~=~$F_{0}$($r$/$r_{0}$)$^{-2}$ relation.
The data roughly follow such a law. In case this emission is due to
recombination and since the ionization flux is decreasing with
distance as $r^{-2}$ this would mean that the number of ionized
hydrogen atoms intersected by the slit is the same at all distances.
This could be the case if the gas is clumpy so that changes in
density has little effect.

\begin{figure*}
\centering
\begin{tabular}{c c}
\includegraphics[bb=107 359 458 685,clip=,width=0.45\hsize]{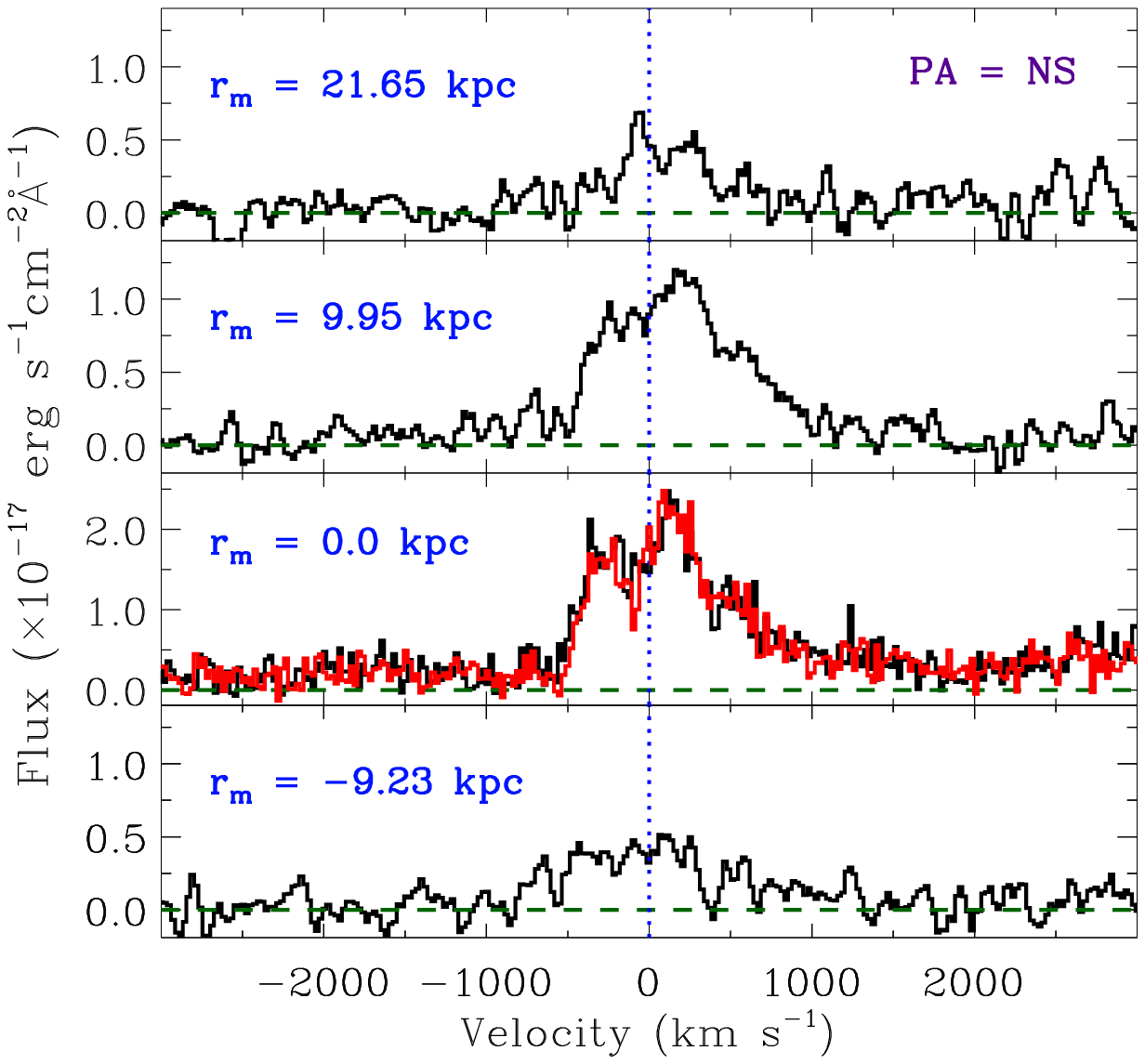}
\includegraphics[bb=107 359 458 685,clip=,width=0.45\hsize]{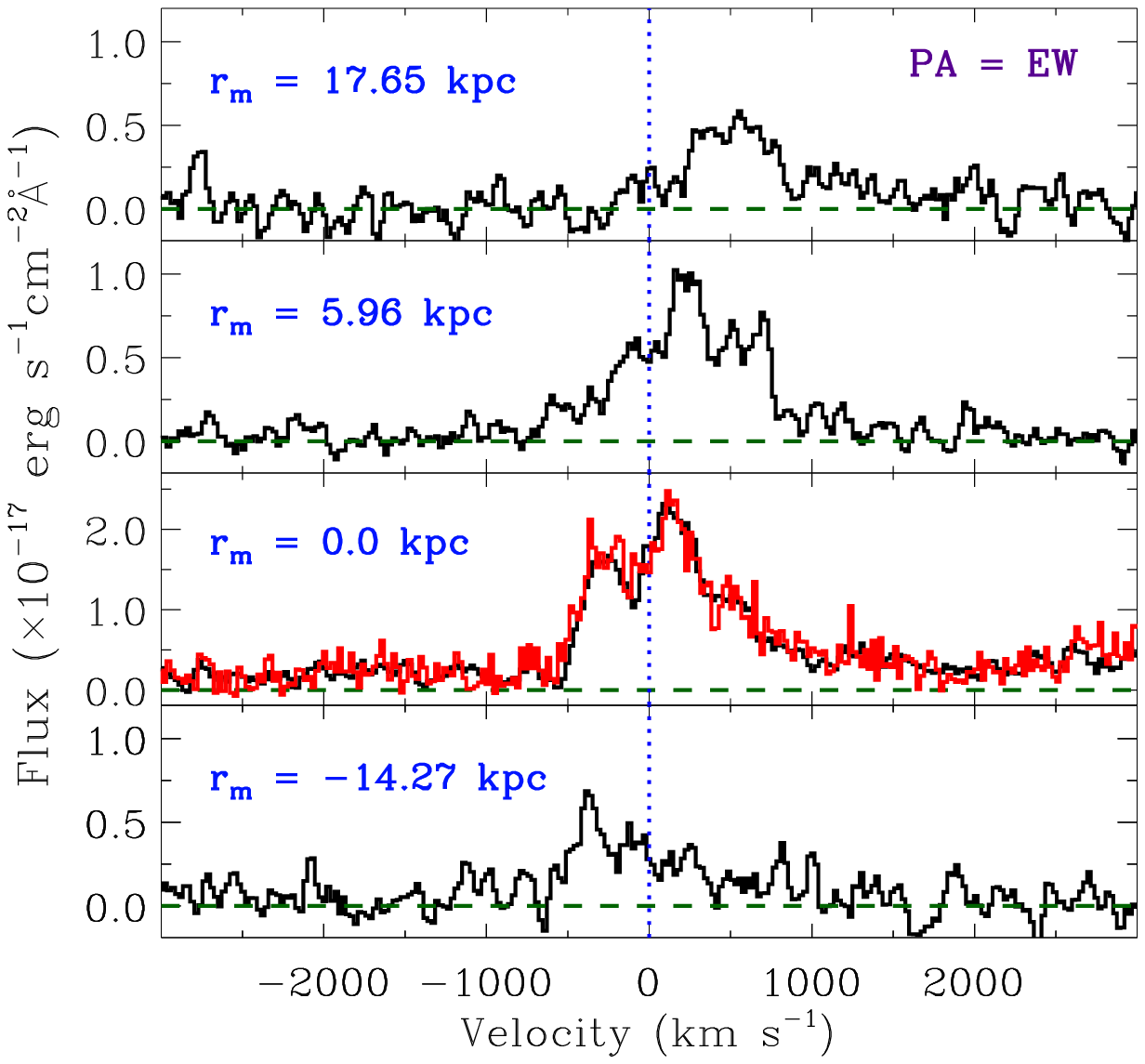}
\end{tabular}
\caption{Velocity profiles of the extended Lyman-$\alpha$ emission
at different distances from the AGN for PA~=~NS (left panels) and
PA~=~EW (right panels). The parameter $r_{\rm m}$ gives the mean
distance to the AGN. The red histogram in the left panel (resp.
right panel) shows the spectrum of the trace for the PA~=~EW (resp.
PA~=~NS).}
 \label{4_Extession_profiles}
\end{figure*}

The velocity profiles of the extended Lyman-$\alpha$ emission at
different distances from the AGN are shown in
Fig~\ref{4_Extession_profiles} and Table~\ref{flux_luminosity}
summarizes the integrated flux and luminosity of the trace and the
extension.

\section{Discussion and conclusions}

We have performed slit spectroscopic observations of QSO J0823+0529
with the MagE spectrograph mounted on the Magellan telescope along
two PAs, in the North-South and East-West directions. The quasar is
unique because a DLA is located at a redshift very similar to that
of the quasar ($z_{\rm DLA}$~=~3.1910 and $z_{\rm CIV}$~=~3.1875,
i.e. $\Delta v$~$\sim$330~km~s$^{-1}$) and acts as a coronagraph
blocking most of the flux from the central regions of the AGN. In
the present case, the DLA cloud is small enough so that it covers
only approximately 90\% of the Lyman-$\alpha$ BLR. This puts us in a
unique position to be able to directly observe the quasar narrow
line region and the extended emission line region. Indeed,
Lyman-$\alpha$ emission is detected up to more than 25~kpc from the
quasar along both PAs.

\subsection{The quasar NLR}

Finley et al. (2013) have gathered a sample of 57 DLAs acting as a
coronagraph in front of quasars; the DLAs have redshifts within
1500~km~s$^{-1}$ from the quasar redshift. Their statistical sample
of 31 quasars shows an excess of such DLAs compared to what is
expected from the distribution of intervening DLAs. This can be
explained if most of these DLAs are part of galaxies clustering
around the quasar. However 25\% of such DLAs reveal Lyman-$\alpha$
emission probably from the quasar host galaxy implying that these
DLAs have sizes smaller than the quasar emission region. The
Lyman-$\alpha$ luminosities are consistent with those of
Lyman-$\alpha$ emitters in 75\% of the cases and 25\% have much
higher luminosities. The later systems probably reveal the central
NLR of the quasar. QSO~J0823+0529 is part of this later class of
coronagraph DLAs.

We have shown that the size of the neutral phase is of order 2-9~pc
which means that the gas does not cover the Lyman-$\alpha$ NLR.
Three main components of Lyman-$\alpha$ emission from the NLR are
clearly detected at $\Delta V$~=~$-$300, +100 and +400~km~s$^{-1}$
relative to the DLA redshift (see Fig.~9). The corresponding C~{\sc
iv} emission seems to be absent in the second strongest component.
Either the Lyman-$\alpha$ emission is scattered or the associated
C~{\sc iv} emission is absorbed by the high-ionization phase of the
DLA. To be covered, the corresponding gas must be located within
$\sim$3-80~pc from the quasar which, we have seen, corresponds to
the extension of the high-ionization phase of the cloud assuming
spherical geometry. At least part of the two other components could
therefore be located beyond this distance from the quasar, although
part of the third component could also be hidden, as the C~{\sc
iv}/H~{\sc i} ratio is larger by a factor of about two in the first
component compared to the third component (see Fig. 9). Note that
the relative velocities of components 1 and 3 can be interpreted as
the result of a conical outflow with mean projected velocities of
about $\pm$300~km~s$^{-1}$.

\subsection{Lyman-$\alpha$ emission from the center to the outskirts of the host galaxy}

Extended Lyman-$\alpha$ emission has been observed around
high-redshift  radio  galaxies (Heckman et al. 1991;  van Ojik et
al. 1996), as well as around radio-quiet quasars (Bunker et al.
2003; Christensen et al. 2006). For radio-loud quasars or
radio-galaxies, the Lyman-$\alpha$ flux of the nebula is an order of
magnitude higher (Christensen et al. 2006), presumably because the
emission of radio-loud quasar gaseous envelopes is enhanced by
interactions with the radio jets. More recently, North et al. (2012)
have used a careful treatment of the spectral PSF to extract quasar
traces. This revealed four detections of extended emission out of
six radio-quiet quasars at $z\sim 4.5$ with extensions of diameter
$16<d<64$~kpc down to a luminosity of
2$\times$10$^{-17}$~erg~s$^{-1}$~cm$^{-2}$~arcsec$^{-2}$. The
emission has 900~$<$~FWHM~$<$~2200~km~s$^{-1}$. Our observations are
in line with these numbers. The extended emission we detect in QSO
J0823+0529 has a diameter of $\sim$50~kpc and
FWHM~$\sim$~900~km~s$^{-1}$. However, QSO J0823+0529 does not seem
to follow the $L$(Lyman-$\alpha$) vs $L$(BLR) relation indicated by
these authors. Indeed the Lyman-$\alpha$ luminosity in QSO
J0823+0529 is more than an order of magnitude larger than what would
be expected from this relation even after correcting for dust
attenuation. It is still possible that dust is present closer to the
quasar and further attenuates the BLR Lyman-$\alpha$ emission. It is
also possible that the decomposition between the broad and narrow
line region emissions was ambiguous in previous studies so that the
NLR emission could have been underestimated.

 \begin{table}
 \centering
\caption{ Lyman-$\alpha$ integrated fluxes and luminosities in the
PSF centered on the quasar trace and in the extension. The last row
gives the integrated flux and luminosity of the broad Lyman-$\alpha$
emission calculated using the simulated spectrum shown in
Fig~\ref{simulateQSO}.}
 \setlength{\tabcolsep}{10.0pt}
\renewcommand{\arraystretch}{1.2}
\begin{tabular}{c c  c c}
\hline

 $ $   &  Integrated~flux             &  Luminosity           \\
 $ $   &  [erg~s$^{-1}$~cm$^{-2}$]    &  [erg~s$^{-1}$]    \\
\hline

Trace(PA=NS) & 3.59~$\times$~10$^{-16}$ &  3.43~$\times$~10$^{43}$   \\
Extension(PA=NS) & 3.97~$\times$~10$^{-16}$ &  3.79~$\times$~10$^{43}$ \\
\hline
Trace(PA=EW) & 3.36~$\times$~10$^{-16}$ &  3.21~$\times$~10$^{43}$   \\
Extension(PA=EW) & 2.68~$\times$~10$^{-16}$ &  2.56~$\times$~10$^{43}$ \\
\hline
Lyman-$\alpha$(BELR) & 7.10~$\times$~10$^{-15}$ &  6.78~$\times$~10$^{44}$   \\
\hline
\end{tabular}
 \label{flux_luminosity}
 \renewcommand{\footnoterule}{}
\end{table}

The emission is more extended to the North-West of the object (see
Figs.~11, 13 and 15). There are two notable features in the spatial
and velocity structure of the nebula. First, in the North-West
direction, the kinematics are strikingly similar along the trace and
10~kpc away from the center (see Fig.~13) with a velocity spread of
more than 1000~km~s$^{-1}$. The emission is quite strong in this
region. Secondly, there is a clear gradient of about
1000~km~s$^{-1}$ between 15~kpc to the East and 20~kpc to the West.
It is tempting to interpret these features as the superposition of
the emission of gas in the disk of the galaxy, where the density is
higher and turbulent kinematics prevent gas clouds to be well
organized with emission from gas flowing out of the disk with
velocities of the order of 500~km~s$^{-1}$. This gas is best seen up
to 20~kpc to the West and 10~kpc to the East. Such winds can be
reproduced by recent models taking  into account  the  effects of
radiation trapping (Ishibashi \& Fabian 2015).

\subsection{Nature of the DLA}
It is well demonstrated that bright quasars are surrounded by large
amounts of gas both in extended ionized halos up to 10~Mpc from the
quasar (e.g. Rollinde et al. 2005) and extended (300~kpc) reservoirs
of neutral gas in the halo of the host galaxy (Prochaska et  al.
2013,  2014; Johnson et al. 2015). However, along the line of sight
to quasars, the incidence of neutral gas is less (e.g. Shen \&
M\'enard 2012) indicating that the  ionizing  emission  from quasars
is highly anisotropic. The DLA we discuss here is part of the
so-called proximate DLAs in the sense that the absorption redshift
is similar to that of the quasar ($z_{\rm abs}$~$\sim$~z$_{\rm em}$,
see e.g. Ellison et al. 2010). However, this is the first time it
can be demonstrated that the gas is located very close to the quasar
($<$~400~pc) and probably associated with the central part of the
host galaxy.

We have modelled the physical state of the gas in the DLA using
photo-ionization models. The ionization parameter is found to be in
the range $-1.1$~$<$~log$U$~$<$~$+0.3$ which means that the cloud is
mostly highly ionized. A density of $n_{\rm
H}$~$\sim$~710-180~cm$^{-3}$ is needed to explain the absorption
lines from O~{\sc i} and Si~{\sc ii} ground state excited levels
implying that the neutral phase should have dimensions of the order
of 2 to 9~pc and be embedded in an ionized cloud of size
$\sim$3-80~pc. From this, we could derive that the cloud is located
between 230 and 580~pc from the quasar. The metallicity of the
cloud, $Z$~=~0.16~$Z_{\odot}$, is typical of the metallicity of
standard intervening DLAs (Rafelski et al. 2012).

It is intriguing to note that the high-ionization phase of the cloud
we see has characteristics similar to those of some of the warm
absorbers seen in many AGNs. Tombesi et al. (2013) argued that warm
absorbers (WA) and ultra-fast outflows (UFO) could represent parts
of a single large-scale stratified outflow observed at different
locations from the black hole. The UFOs are likely launched from the
inner accretion disc and the WAs at larger distances, such as the
outer disc and/or torus. There are still significant uncertainties
on the exact location of this material, which ranges from a few pc
up to kpc scales, (e.g. Krolik \& Kriss 2001; Blustin et al. 2005).
The absorption lines are systematically blue-shifted, indicating
outflow velocities of the WAs in the range 100-1000 km~s$^{-1}$.
 King \& Pounds (2013) have argued that the dense gas which
surrounds the AGN when it starts shining is swept out by the fast
winds powered by the accretion luminosity. The wind is halted by
collisions near the radius where radiation pressure drops. The
shocked gas must rapidly cool and mix with the swept-up ISM.
Distance from the AGN and properties of the gas are similar to what
is observed for warm-absorbers, as in our case. Therefore, in
QSO~J0843+0529, the DLA could be located in the galactic disk at the
terminal position of the wind, at the limit of the interstellar
medium of the host galaxy.  The presence of dust in the gas can be
considered as supporting this view as it is expected in the dense
environment of AGNs (see e.g. Leighly et al. 2015).

 The only caveat with this idea is that the metallicity of the
DLA is typical of intervening DLAs, when we would expect the gas in
the ISM of the quasar host galaxy to have larger metallicity. This,
together with the low outflow velocity (the DLA is centered on the
Lyman-$\alpha$ emission), argues for another explanation for the
origin of this gas. Given the distance to the central AGN
(230~$<$~$r_0$~$<$~580~pc), one is tempted to conjecture that it
could be the final fate of infalling gas. Note that in that case,
the presence of dust is not a problem as the observed amount is not
unusual compared to what is seen in typical high column density
DLAs. Indeed, accretion is believed to happen through cold flows
which are expected to be of lower metallicity compared to the
environment of the quasar (see Bouch\'e et al. 2013). Little is
known about how quasars at high redshift interact with cold
infalling streams of gas, and in particular whether these collimated
structures can survive the energy released by the AGN. Therefore the
DLA cloud could be in a transitory phase before it is completely
destroyed by the AGN ionizing radiation field (Namekata et al.
2014).

The two alternatives are actually not incompatible, since some of
the warm absorbers could be part of the galactic ISM swept away from
the center by the AGN and placed at large distances from the AGN
(e.g. King \& Pounds 2013). These outflowing warm absorbers may then
intersect with accreted cold streams.

\subsection{Concluding remark}
There is little doubt that DLAs acting as coronagraphs are important
targets to be observed and analyzed. Besides revealing interesting
characteristics of the quasar host galaxy they are, like in QSO
J0823+0529, potentially part of the machinery that power the AGN.
Finley et al. (2013) found 57 such systems out of about a third of
the BOSS targets. With the on-going eBOSS-SDSSIV survey and the
forseen DESI survey (Schlegel et al. 2011), the number of such
systems will increase by a large factor in the future. It will thus
be important to gather a large sample of these objects to be able to
study their characteristics statistically.

\section*{Acknowledgments}
We would like to thank the anonymous referee for their constructive
comments, which helped us to improve the paper. We also thank George
Becker for advices on MagE data reduction and Hadi Rahmani for
useful discussion. H.F. was supported by the Agence Nationale pour
la Recherche under program ANR-10-BLAN-0510-01-02. SL has been
supported by FONDECYT grant number 1140838 and partially by PFB-06
CATA.

\end{document}